# An Adaptive E-Learning System Using Justification Based Truth Maintenance System

Tahir Mohammad Ali [1], Attique Ur Rehman [2], Ali Nawaz[2], Wasi Haider Butt[2].

[1]Department of Computer Science, Gulf University of Science and Technology, Kuwait
[2]Department of Computer and Software Engineering, CEME, NUST, Islamabad, Pakistan
Corresponding author: Tahir M. Ali (e-mail: ali.t@gust.edu.kw).

*Abstract*- In most E-learning systems, educational activities are presented in a static way without bearing in mind the particulars or student levels and skills. Personalization and adaptation of an E-learning management system are dependent on the flexibility of the system in providing different learning and content models to individual students based on their characteristics. In this paper, we suggest an Adaptive E-learning system which is providing adaptability with support of justification-based truth maintenance system. The system is accomplished of signifying students with suitable knowledge fillings and customized learning paths based on the student's profile, interests, and previous results. The validation of proposed framework is performed by meta-model.

*Index Terms*-- Adaptive E-Learning Management System, Justification Based Truth Maintenance System, Decision Support Systems.

## I. INTRODUCTION

Education plays important rule in the economic, social and overall development of country. Education is not only an essential tool to improve the life standard of individual but also improve the living standard of society. Meanwhile, the biggest challenge we face as a society is providing the quality education which is according to the dynamics and intellects of individual learners. With the advancement in the field of computer technology application of computer science in all field including education tremendously rises. Especially in situation like COVID-19 where everyone is restricted to limited space the relying on technology and communication is increases therefore it is important to develop the systems which cater all learning requirements. Learning management systems (LMS) are prevailing unified systems that provide a number of events that can be accomplished by instructors and learners during the e-learning process [1]. Instructors mainly use an LMS to prepare and deliver course materials, quizzes, tests, connect with students and evaluate student progress. In typical MS the same learning assets are provided to different learners. Learners have different types of abilities, preferences, and numerous kinds of content on knowledge subjects are essential to familiarize the knowledge content to the necessities of diverse learners. Adaptive learning is a serious prerequisite for e-learning systems which vigorously adapts learning content to the student educational needs for endorsing learning results [2].

Therefore, in this paper, we are proposing an Adaptive E-learning management system constructed on Justification based Truth Maintenance System. The contribution of the paper are summarized as following;
- Proposed an adaptive e-learning framework based on truth maintenance system
- Validation of proposed framework is performed by building meta-model

Our proposed system is proficient of signifying students with suitable learning contents and customized learning paths based on the student's profile, interests, and previous results.

The rest of the paper is organized as follow: Section II describes the literature review; the problem statement and proposed framework is discussed in Section III and Section IV conclude the paper.

## II. LITERATURE REVIEW

Personalization and adaptation of an LMS is dependent on the flexibility of the system in providing different learning and content models to individual students based on their characteristics. The idea of adaptation has been a significant issue of research for learning systems in the last few years. Different artificial intelligence and data mining techniques and algorithms were used to propose solutions for adaptive learning paths for the-learning management system. Authors in [3] present a model to create a knowledge strategy recommendation system created on graph theory. The approach uses randomly distributed data set competencies among the learning objects. A system based on the data mining technique to provide modified learning paths for enhancing the performance of imagination is proposed in [4]. A physical model recommender system framework based on successive pattern mining is presented in [5] and multidimensional attribute-based cooperative clarifying. From [6], presented an approach for recommending suitable learning paths using the swarm intelligence model [7-16] proposed a system for obtaining learning paths as a restraint fulfilment problem in which meta-data and capabilities are used to define the associations between the learning objects, where the course materials are used to formulate LOs sequence.

## III. RESEARCH PROBLEM AND PROPOSED SOLUTION

The core objective of this study is to develop a system that can suggest adaptive knowledge paths built on prior knowledge information about students. It suggests adaptive learning pathways to learners based on their previous learning activities and results, such as, asking the learner to repeat a prerequisite topic or read more examples, repeat a topic in more detail, taking extra trainings with high or fewer difficulty levels. Personalization and adaptation of an LMS is dependent on the flexibility of the system in providing different learning and content models to individual students based on their characteristics.

### A. Adaptive E-Learning System Using The Justification Based Truth Maintenance System

In this research, we are going to propose an Adaptive E-Learning system using the Justification Truth Maintenance System, which is accomplished of recommending students with suitable learning content and customized learning paths based on the student's profile, interests, and previous results.

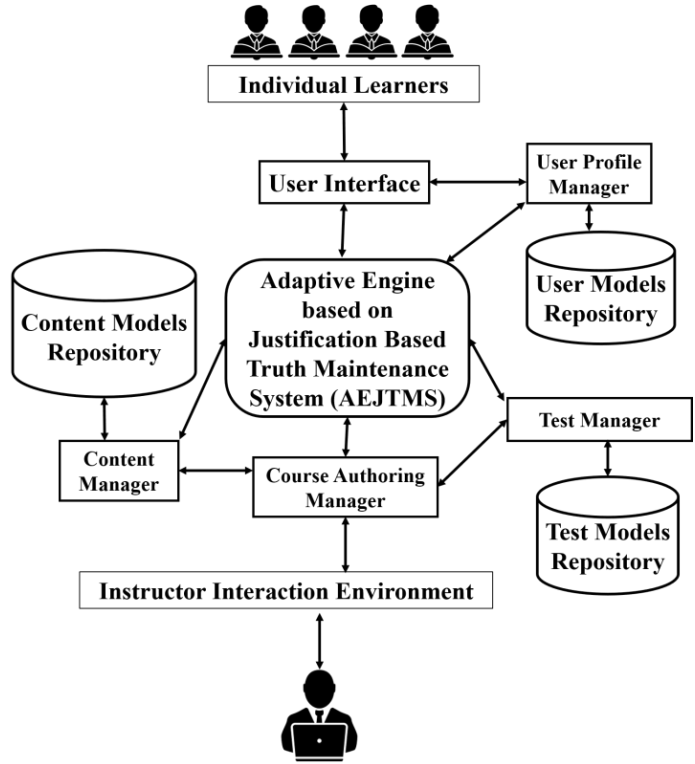

FIGURE 1: Proposed Framework of Adaptive E-learning with Justification based Truth Main

The proposed architecture of the system is depicted in Fig. 1. The proposed system involves of mainly six components: User Interface, Content Manager, Course Authoring Manager, User Profile Manager, Test Manager and the core module which is the Adaptation Engine based on the Justification based Truth Maintenance System. User Interface module deals with the student logins, initial registration and providing the customized learning paths recommended to the learner by the Adaptation Engine. Course Authoring Manager module permits the trainers to create or inform a course gratified and test repositories. The other modules are responsible for storing and providing the required information to the core module which is the adaptation engine that is accountable for providing adaptive learning paths as per the student's features or based on the previous test or exercise results of the learner.

### B. The Role Of Jtms In Supporting The Adaptive Learning Paths

The system tracks the course syllabus items as a set of connected nodes in the JTMS network. A node in the JTMS network represents a single knowledge milestone that is required to be achieved by the learner as a part of the course curriculum. A node is attached with a set of learning assets that allows the student to grasp and achieve the target knowledge of the node. Also, the node is linked to those activities that will examine the student mastering level of the target node knowledge. The in-list of the JTMS node is representing the list of prerequisites nodes that needs to be completed by the learner before exploring the current node activities. The out-list of the JTMS node is representing the list of nodes where this node is prerequisite.

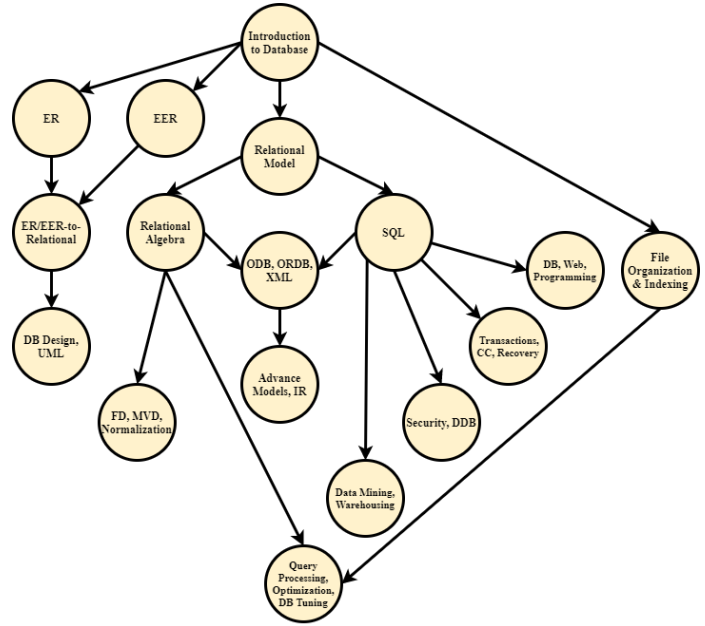

FIGURE 2. A high level JTMS network for a typical database course

Figure 2 represents a high level JTMS network for a typical database course. A copy of the JTMS network is assigned to each student when the student is enrolled in the course. The teacher is allowed two options for the initial setup:

Allow the students to explore all nodes assets in any order this is similar to the textbook concept. Students are able to explore the node assets only when all the nodes in the in-list are marked as passed (see Fig. 3).

TABLE 1: JTMS Node Status/Coloring Scheme.

| Status | Color |
|---|---|
| Locked | Red |
| Exploring | Yellow |
| Passed | Green |

45

For each node the system tracks the student status. Table 1 summarizes the JTMS Node Status Scheme. Table 2 provides the student mastering level scheme for each node. Tracking the student master level will help the system to provide adaptive assets and recommendations for each student as we will discover in the next section.

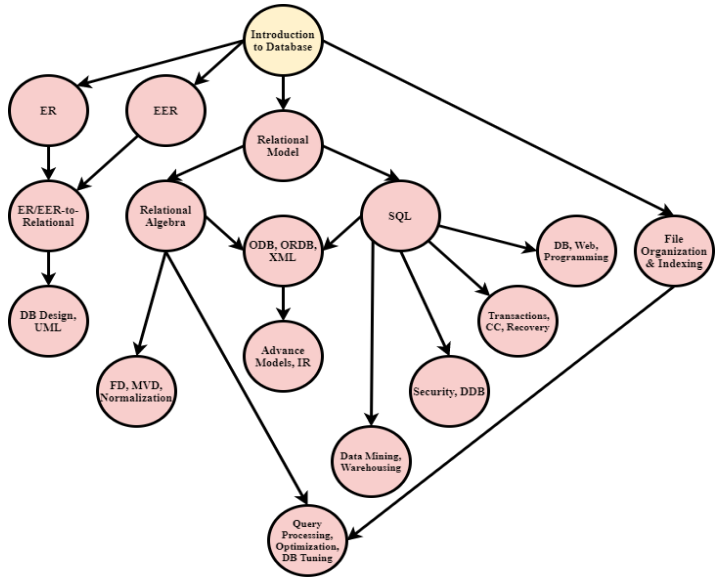

FIGURE 3. A high level JTMS network for typical database course with locked nodes.

TABLE 2: Node mastering level.

| Level | Description |
|-------|-------------|
| 1 | Minimum |
| 2 | Average |
| 3 | High |
| 4 | Excellent |

C. ADAPTIVITY DISCUSSION

The system can provide adaptability in E-learning management system from different aspects:

- Tracking the status of each student in covering the course content and the master level of each knowledge skill in the course. Figure 4 are demonstrating the JTMS network for two different students in the course.
- Tracking the master level of each node helps the teachers to provide different learning assets based on students' achievements. e.g., for those who are doing well in the course; the system allows them to advance and challenging exercises. In contrast, students who are suffering in the course can be supported by providing them extra material that helps them to understand the topic. Furthermore, when a student gets stuck in a certain topic, the system can recommend the student to revise the prerequisite of the topic based on the performance of the student in the prerequisite nodes.

Referring to Fig. 5, let's assume the student is struggling in Module (node) "ODB, ORDB, XML" of the course, the system may recommend the student to revise the prerequisite node "Relational Algebra" first since the student level is grasping the node knowledge was not very strong. If the problem still remains, then the system recommends to revise the other prerequisite node "SQL" and so on.

- Displaying the material based on preferred learning styles of the students. For example, some learners prefer text-based material, others may prefer video lessons, a third group of learners may prefer interactive lessons. Note that this feature is not incorporated in the current design of our system however it can be added.

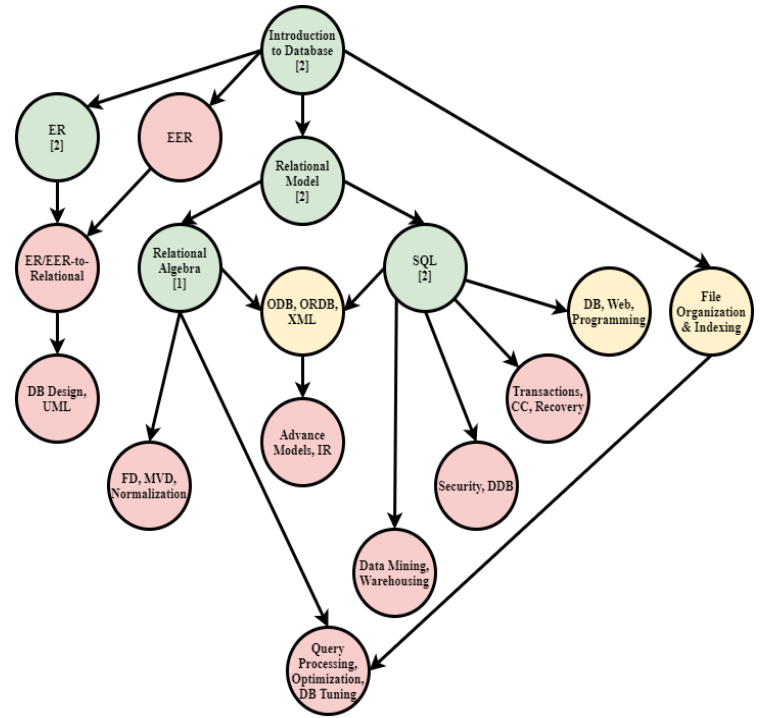

FIGURE 4. A JTMS network for Student 52.

IV. VALIDATION

The validation of proposed framework is performed eclipse modeling framework. The Eclipse Modeling Framework (EMF) is a set of Eclipse plug-ins which can be used to model a data model and to produced code or other output based on this model. The meta-model is consisting of eight classes. All the seven shown in Fig. 5, send the information to Adaptive Engine based Justification Truth Maintenance System (AJTMS) class. AJTMS is a final class that is responsible providing recommendation to the students.

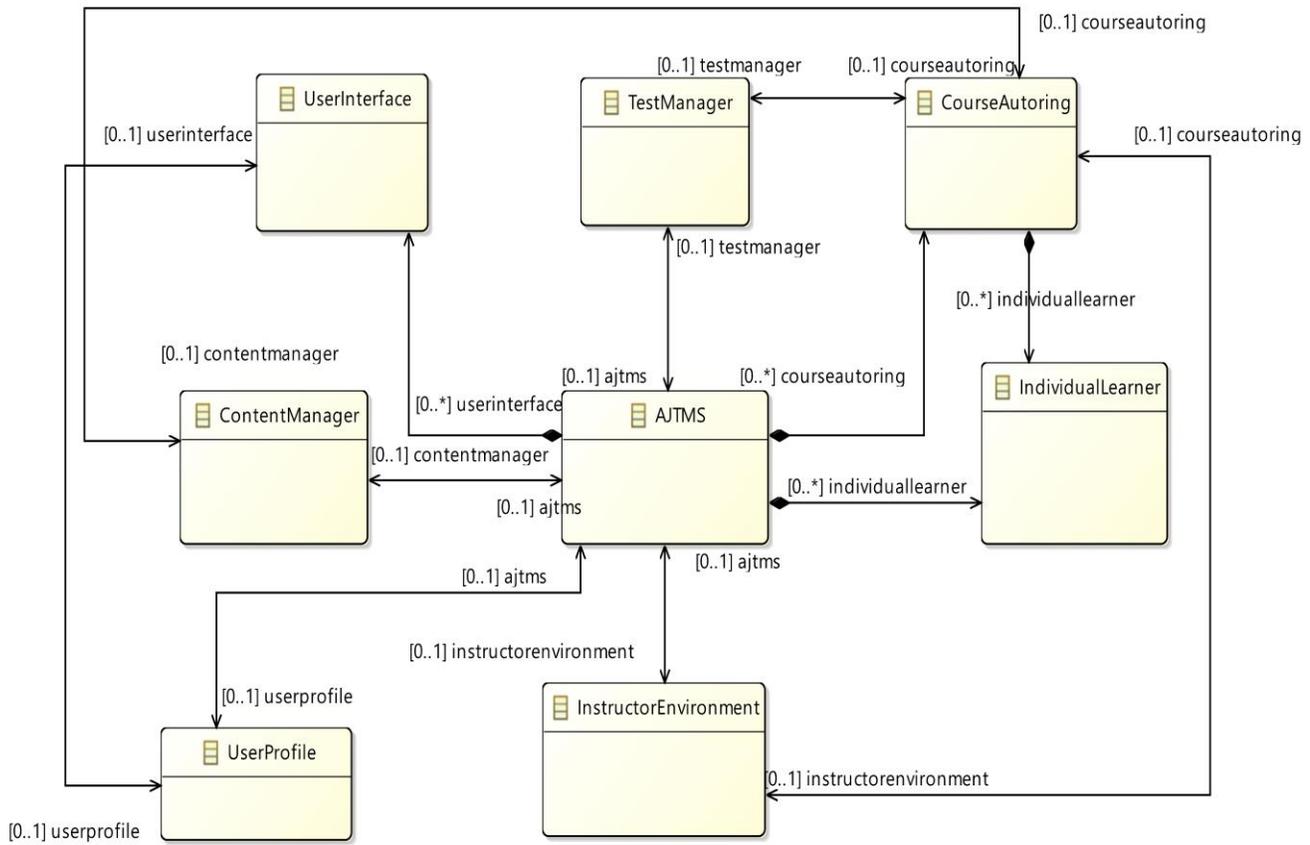

FIGURE 5: Meta-Model of Proposed Framework

## V. CONCLUSION

In this paper, a new adaptive knowledge path support for e-learning management scheme is anticipated. Our approach is based on the support of justification-based truth maintenance system. The system can provide adaptability in E-learning management systems from different aspects. Our proposed solution is supporting the natural idea of a knowledge map where each node is linked to its prerequisites and post requisites. The JTMS network propagates the student progress through the network and behaves according to the student's level, preferences and capabilities.